# Symbiotic Optimization of the Nanolithography and RF-Plasma Etching for Fabricating High-Quality Light-Sensitive Superconductors on the 50 nm Scale

H Bartolf[1*], K Inderbitzin[1], L B Gómez[1], A Engel[1], A Schilling[1], H-W Hübers[2,3], A Semenov[2], K. Il'in[4], M. Siegel[4]

[1]Physics-Institute, University of Zürich, Winterthurerstrasse 190, 8057 Zürich, Switzerland

[2]Institute of Planetary Research, German Aerospace Center (DLR), Rutherfordstrasse 2, 12489 Berlin, Germany

[3]Department of Optics and Atomic Physics, Technical University Berlin, Hardenbergstrasse 36, 10623 Berlin, Germany

[4]Institute for Micro- and Nano-Electronic Systems, Karlsruhe Institute of Technology, Hertzstrasse 16, 76187 Karlsruhe, Germany

*E-mail: holger.bartolf@physik.uzh.ch, +41 44 635 5755

**Abstract** We present results of a fabrication-process development for the lithographic pattern transfer into the $sub-100nm$ range by combining electron-beam lithography and reactive dry etching to obtain high quality niobium-based light-sensitive superconducting devices. To achieve this spatial resolution, we systematically investigated the stability of the positive organic etching masks ZEP 520A and PMMA 950k in different properly operated fluoride-based plasma discharges. The chemically more robust ZEP 520A was used for defining the nanoscaled superconductors during the dry plasma etching. Our etching recipe is appropriate for a precisely controlled removal of a number of transition metals, their nitrides and a number of lithographic resists. Our process yielded light-sensitive superconducting devices made from NbN with smallest planar lateral dimensions of about $50nm$ with a critical temperature $T_c(0)$ of about $13K$, which is close to the transition temperature of the unstructured thin film. Our ultra-narrow current paths are able to permanently carry bias-currents up to 60% of the theoretical de-pairing current-limit.

1. Introduction

Research and development (R&D) on superconducting micro- and nanostructures is driven by their immense potential for applications as thermal [1,2] or ionizing [3,4] single-photon cryogenic radiation- and/or particle-detectors [5-7], Josephson-junctions [8], **S**uperconducting **QU**antum **I**nterference **D**evices (SQUID's) [9], quantum-cryptography [10-12] and quantum-information [13-15] devices. As compared to those devices, a superconducting nanowire highspeed single-photon detector [16,17,18,19] (see figure 1, where an electrograph of such a device is shown) for the visible and near-infrared spectral range requires smallest planar device dimensions and hence a well thought-out fabrication procedure. At spatial planar widths $w$ of the nanowire of about 50nm, which corresponds to only a few hundred atoms in that dimension, standard top-down fabrication [20] methods meet the limits where bottom-up methods [21] dominate. Based on fundamental arguments of statistical physics, devices fabricated on these ultra-narrow nanoscales exhibit a pronounced sensitivity to absorbed photon energies. As a general rule of thumb, the sensitivity of the detector increases as the detection elements volume decreases. At the same time, the device becomes more susceptible to thermal and/or quantum-mechanical fluctuations [19] that manifest themselves in a higher noise of the operational detector; a physical frontier that also confronts manufactures of modern digital cameras ([22], micron-scaled pixels), where the effect is even more pronounced because the strength of room temperature thermal fluctuations is about two orders of magnitude higher, as compared to a device built from a low-temperature superconductor.

In general, top-down fabrication technology on the nanoscale makes it possible to design superconductors with spatial variations on the order of the microscopic characteristic length scales of the superconducting state, such as the magnetic penetration depth $\lambda$, the coherence length $\xi$ or the inter-vortex spacing [23]. In these reduced spatial dimensions, finite-size effects may dominate the physical properties of the superconducting state and manifest themselves, for example, in a temperature-induced rounding of the electronic phase transition depending on the underlying fluctuation mechanism that supplies fluctuating energy into the electronic system (see the introductorily chapter in [24] and reference [25] for a detailed discussion). Well below the phase transition temperature $T_c(0)$, fluctuations provide sufficient energy to excite paired vortices in the electronic system even without an externally applied magnetic field, which manifests itself in the well-known Berezinskii-Kosterlitz-Thouless (BKT) phase transition [26,27]. Characteristic temperature dependencies of experimentally measurable quantities (see e.g. Fig. 2. in [19]) reflect the physical nature of the underlying fluctuation mechanism inducing the transition. Therefore, the development of lithographic fabrication technology in the $sub-50\text{nm}$ regime opens up new windows for raising and answering fundamental questions concerning superconductivity [19]. To explore these questions experimentally and to explain them quantitatively within the established frameworks of the seminal quantum-statistical theories describing the superconducting and the metallic state, it is necessary to develop lithographic fabrication procedures of very high quality and repeatability.

The commonly employed technological approach for top-down fabrication is electron-beam lithography (EBL) [20,28,29] that enables the lithographic pattern transfer well below the $20\text{nm}$ length scale [30-32] even at electron accelerating voltages of $30\text{kV}$ (R&D relies on a relatively low cost accessory, so that we restrict the discussion here to such systems). This relatively high acceleration voltage is accompanied by a pronounced lithographic proximity exposure which can be intrinsically suppressed for planar device layouts that are orders of magnitude narrower than the backscattering length of the electrons (see chapter 2 of reference [20] for a detailed discussion), which is true for fabrication length scales on the $sub-100\text{nm}$ scale. However the fabrication in this regime is challenging in general in case of superconductors because the exposure of tiny metallic structures to air at room temperature after the successful fabrication procedure generates a nanometer-thick normal-conducting cloak around the superconducting core (refer to the elaborated work of J. Halbritter [33] and the references therein in case of Nb-based structures) that weakens the superconductive state and hence lowers the transition temperature and the experimentally observable critical-current due to the superconducting proximity-effect [34]. The influence of this effect becomes more pronounced for smaller lateral dimensions of the superconductor and might even lead to the total suppression of the superconducting state. Therefore fabri-cation procedures that have a negligible deteriorating impact on the superconductive properties of the device are desirable.

The lithographic pattern transfer is accompanied by a certain planar line-edge roughness [35] at the vertical metal-vacuum interface, i.e. at the edges of the latter nanoscaled structured superconductor. The absolute value of this lithographically generated border-line depends on the resist's contrast and on the nature of the lithographic pattern transfer (the RF-plasma conditions in case of a dry etching approach). As a rule of thumb, it has to be assured that the width of this crossover-borderline is at least one order of magnitude smaller than the narrowest planar part of the structure to eliminate a drawback of this fabrication uncertainty on the physical properties of the device. This can be achieved by using thinner resists that increase the resolution during the EBL pattern generation (see figure 1A in [31]) and by a low-temperature chemical development of the exposed electron-sensitive resist that increases the contrast of the resist and therefore the achievable resolution [36]. In addition, a relatively small electron-beam current during the EBL eliminates dynamic positioning uncertainties of the beam occurring during the line-scan that defines the latter metal-vacuum interface. Due to the fact that the small beam-current is accompanied by a relatively long exposure time, a mix and match of photolithography and EBL is favourable (refer to chapter 4 in [20]). Finally, the proper operation of the reactive plasma-discharge ensures a sensitive ablation of the unprotected superconducting film, controlled with high precision during the dry lithographic pattern transfer, which conserves the high resolution of EBL. It is the uniqueness and novelty of this work to discuss the symbiotic parameter optimization during the subtractive pattern transfer using EBL and subsequent reactive dry-etching to obtain ultra-narrow (50nm) superconducting paths from NbN with a critical temperature $T_c(0) \cong 12.6K$ which is close to that of the unstructured film $\cong 13.2K$.

This paper is organized as follows: We first describe the preparation of the thin superconducting films. Afterwards we present the development of a smooth, controlled and selective ablation of the generated films by properly operating a reactive plasma discharge, and we discuss the determined etching rates for several superconductors and organic resists which let us decide in favour of the proper etch mask material to obtain a thin etch mask and hence a sub−100nm resolution according to figure 1A in [31]. The following chapter describes the necessary EBL considerations that allow for a high resolution, planar top-down pattern transfer into the 50nm regime (see figure 1). Finally we discuss the electronic transport measurements on the obtained light-sensitive (see chapter 13 in [18]) meander-detector structures that are able to permanently carry bias-currents up to 60% of the theoretical de-pairing current-limit.

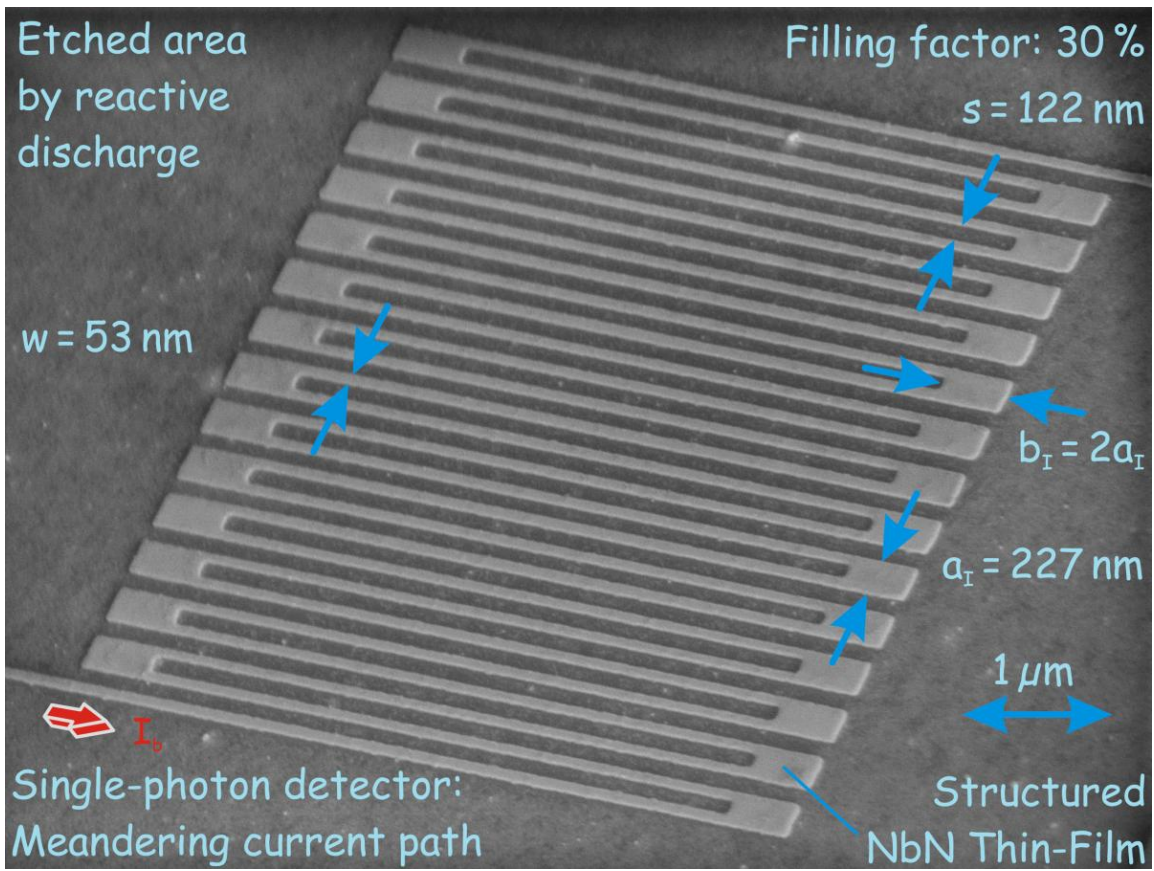

Figure 1: Tilted electrograph of a device clone of a nanoscaled single-photon meander detector which consists of a meandering superconducting path. The squared detection area $A_{\text{detec}}$ covers about $25 \mu m^2$. After EBL, the structure was protected while the reactive discharge etched 8nm

into a 10-nm-tick NbN thin-film sputtered onto an Al$_2$O$_3$ substrate. The remaining 2nm NbN ensures the drain of the electrons from the scanning-beam during the electrography and therefore eliminates charging effects during the scan. The islands (I) that interconnect the conduction paths and that cover an area $a_I = b_I$ were placed to avoid a possible constriction while the conduction path turns around by 180°. $b_I = 2a_I$ was set for scaling purposes.

## 2. Thin-Film Preparation by DC-Magnetron Sputtering

The importance of niobium nitride as a low-temperature superconducting material with a high transition temperature $T_c(B=0) \approx 15\text{K}$ in zero magnetic field $B$, which separates the metallic from the superconducting state, and a large critical transport-current density is well known. These properties were discovered by G. Aschermann *et al.* [37] during experimental investigations of the bulk material. It was later found by T. H. Geballe, B. T. Matthias *et al.* [38] that the transition temperature raises with increasing nitrogen content. D. Gerstenberg and P. M. Hall [39] were among the first to study thin NbN films. The bulk material NbN is a distinct type-II superconductor in the dirty limit [23] that exhibits a very high upper critical field $B_{c2}(T=0) \cong 20\text{T}$ [40] which results in a coherence length $\xi(0) \cong 4\text{nm}$ of the superconducting condensate that is only a few unit cells wide (see also chapter 5.2 for ultra-thin and ultra-narrow nanostructures).

Our NbN films were deposited by direct current (DC) - reactive magnetron sputtering [41,42] of a pure niobium target in an Ar + N$_2$ gas mixture on top of a sapphire (Al$_2$O$_3$) substrate. The partial pressure of the nitrogen was adjusted to $6 \cdot 10^{-4}\,\text{mbar}$ at a total pressure of about $10^{-3}\,\text{mbar}$. The epipolished *R*-plane sapphire substrates were kept at a temperature of 750°C on the anode during the deposition process. This high thermal energy eases the condensation of the sputtered mesoscopic particles into their thermodynamic equilibrium lattice positions and ensures therefore a homogenously growing epitaxial thin-film (see Fig. 1(a) in [43] for a transmission electron microscope (TEM) picture). A voltage of $V = 317\text{V}$ which is larger than the plasma's breakdown-voltage $V_\text{B}$ (see figure 3 in [41]) under operational conditions was applied to the electrodes and ignites the discharging Townsend's [44] avalanche that ionizes the atomic species within the plasma discharge and leads to a disruptive current of about $145\text{mA}$. The chemical reaction of Nb to NbN occurs during the ionic bombardment at the target surface located at the cathode. The sputtered mesoscopic particles redeposit on the substrate(s) located on the opposite electrode. Magnetic fields generated by permanent magnets [42,45] below the cathode are used to confine the ionized discharged particles in order to increase the ionizing rate and hence the probability for Townsend's avalanche [44]. This procedure allows for a stable operation of the plasma even in the used low-pressure regime which is in general accompanied by a logarithmic divergence of the breakdown voltage (known as Paschen's law [46]).

The thickness $d$ of the resulting films was inferred from the sputtering time and a predetermined deposition rate of $0.17\text{nm/s}$ (obtained from measurements with an atomic force microscope AFM; see Fig. 1(b),(c) in [43]). The film growth was optimized with respect to the total and partial pressures of Ar and $N_2$ and the deposition rate to provide the highest phase transition temperature $T_\text{c}(0)$ for a given film thickness (see Fig. 1. in [47]). Sapphire was chosen as a substrate material because its lattice parameters are closer to those of epitaxially grown NbN than the ones of a silicon wafer, therefore allowing for growing superconducting films of much higher quality [48]. In addition, the sapphire serves as an etch stop layer in contrast to the silicon where an additional layer has to be deposited between the substrate and the NbN to avoid under etching caused by the much larger etching rate of silicon (see figures 4(a) and 5(c)).

A number of additional superconducting transition-metal thin films were grown in the same manner. The discharge parameters for sputtering the other materials were identical to the ones used for growing NbN with the following differences. Pure elemental films (Nb, Ta) were depos-ited in a pure Ar atmosphere ($10^{-3}$ mbar), the other nitrides (TaN, TiN, MoN) in Ar+$N_2$ gas-mixtures similar to NbN. Substrate temperatures during deposition were kept at 500°C to 700°C, except for MoN that was grown with the substrate at room temperature resulting in an amorphous film [49]. All films with thickness $d$ >10nm showed superconductivity below their respective critical temperatures, in good agreement with already published data [40]. These films were used for AFM-etch-rate determinations in order to prove the broad applicability of our dry-etching process to other superconducting materials than NbN (see figure 4(a) and table 1 in section 3.2).

## 3. Plasma Etching

One might conclude that a sputtering technique as described in the last chapter is appropriate to remove parts of the generated films by simply placing them on the cathode. However, the relatively high DC power applied to the electrodes to sustain the plasma discharge, in combination with the high ion acceleration voltage, results in a relatively high energy per ionized atomic particle during the bombardment. This is not suitable for removing the superconducting films with accurate nanoscaled precision (as shown in figure 1). To remove the generated films sensitively, the well known and established alternating current (AC) reactive plasma discharge technique was employed (Plasmalab 76 from *Oxford Instruments*), which is also appropriate to account for the insulating character of the sapphire substrate. In the next section we will discuss why we operated the discharge inside the Plasmalab 76 under minimum power and pressure conditions. In section 3.2 we present the AFM-determined etching and sputtering rates for different metallic, organic and semiconducting materials.

*3.1 Considerations for a Proper Operation of the Discharge*

In general, an AC-powered discharge [50] operates in much the same way as a DC-discharge. The only difference is that the applied voltage oscillates with time, and one might think of the discharge switching on and off at twice the driving frequency of the electric field between the electrodes. Such an oscillating discharge has the advantage that it induces an additional source of ionization other than the secondary electrons ejected from the electrodes during the ionic bombardment [51] that is essential to ignite the Townsend's avalanche [44] in the DC case. This additional ionization source are electrons in the negative glow region (see figure 6 in [52]) that are heated as a consequence of elastic collisions [53] with gas atoms due to oscillations [54] in the applied, harmonically alternating electromagnetic field. Consequently, to gain the maximum potentially available ionization rate, the AC-oscillation frequency generating the plasma has to lie in the radio-frequency (RF) regime (13.56MHz for the used Plasmalab 76). Due to the three orders of magnitude smaller mass of the electron as compared to an ionic mass [55], the plasma frequency for the electrons lies in the GHz regime, while the one for the ions is about three orders of magnitude smaller. Therefore a RF frequency of 13.56MHz allows the electrons to gain resonant power from the oscillating field between the electrodes. At the same time, the ions are not able to follow these oscillations due to their mechanical inertia. This heats up the electrons in the plasma and therefore increases the ionization yield.

As a direct consequence, the damage of the sample surface is reduced because the plasma can be operated at a much lower electrical power which is accompanied by a less kinetic energy per particle during the ionic bombardment. The high voltage that is essential for the generation of secondary electrons when operating a DC-glowing discharge (see chapter 2) is no longer needed here to maintain the RF-glow. The electrons resonantly gain high energies equivalent to tens of thousands of degrees Kelvin, thereby allowing high-temperature type reactions that result in the creation of reactive free radicals even in relatively low temperature chemically inert gases as the ones discussed below. This results in a variety of chemically based mechanisms that can be exploited for accurate thin-film removal if the RF-discharge is operated properly.

The mass difference between electrons and ions also results in a higher mobility of the electrons. Therefore one electrode collects more electrons per half-cycle than the oppositely charged electrode is able to collect ions (see figure 9 in [52]) in the meantime. Consequently, a negative bias voltage $V_{DC}$ develops between the electrodes and the plasma-sheath [56]. Although the ions are not able to follow the RF-oscillations of the electric field, they can enter the sheath due to their thermal Brownian motion [57]. The positively charged ions are then accelerated by $V_{DC}$ towards the electrode and hit the surface of the sample where plasma etching can finally proceed in four different basic mechanisms, namely pure physical sputtering, pure chemical etching, ion-enhanced energetic etching (Reactive Ion Etching RIE) or ion-enhanced inhibitor etching (see figure 1 on page 93 in [58]), depending on the operational discharge conditions. More sophisticated information about plasma processes can be found in [41,52,59], historically interesting reviews in [60].

In the following we assume that under our operational conditions ion-enhanced energetic etching takes place. This type of etching occurs when an ion accelerated through the sheath hits the surface. The released kinetic energy of the charged ion induces the reaction of the sample surface with an electrically neutral, but chemically reactive gas species. Since the ions are accelerated across the plasma sheath and strike the electrode surfaces vertically, the induced etching is directional. At this point the pressure plays a crucial role. If, for example, the pressure is sufficiently high, the ions effectively share the maximum potential energy induced by $V_{DC}$ among several ions and neutrals, and the net bombardment energy per particle [56] decreases. At the same time, the directionality of the incident ions is reduced by scattering effects, and any etching process using these ions becomes less anisotropic. For this reason, the plasma has been operated at the lowest possible pressure $p \approx 10 \mu bar$ that allows for the ignition of the discharge. At low $V_{DC}$ values and hence low operational powers, the discharge has sometimes not been ignitable, however. In this case more power has manually been supplied to ignite the discharge. As soon as the glowing discharge stabilized, the power has then manually been reduced until the desired $V_{DC}$ was reached. An additional parameter that we have not further investigated is the temperature of the electrode on which the substrates were placed and we fixed it to $5°C$.

*3.2 Etching Rate Determination*

The etching rates $\gamma$ for different superconducting materials and organic lithographic resists were finally measured with an AFM (Research AFM from *Asylum*) for the very thin sputtered or spin-coated films $(<50\text{nm})$, and with a standard surface profiler (Alpha-Step 500 from *Tencor*) for thicker films $(>50\text{nm})$. In the latter case, structures prepared by optical lithography [61] were sufficient for this purpose. For the thin NbN-films used in detector applications (see figure 1), the etching rates were determined using a relatively thick $(15\text{nm})$ film that was solely processed for the measurements of the etching rate. In this case the electron-sensitive ZEP 520A [62] (from *Nippon Zeon Co.*) was used as lithographic resist-mask and EBL (see chapter 4) to structure it on the relatively large microscale to ease the use of the AFM. The resist height $H_R$ after the photon (or electron exposure, respectively), and the chemical development was measured before the sample was put inside the plasma reactor chamber, and after the etching step, $H_E$ was determined. Finally, the resist residuals were stripped away in 85°C hot NMP (N-Methyl-2-Pyrrolidone $C_5H_9NO$, an organic solvent) and $H_S$ was then measured. The whole procedure is shown schematically in figure 2.

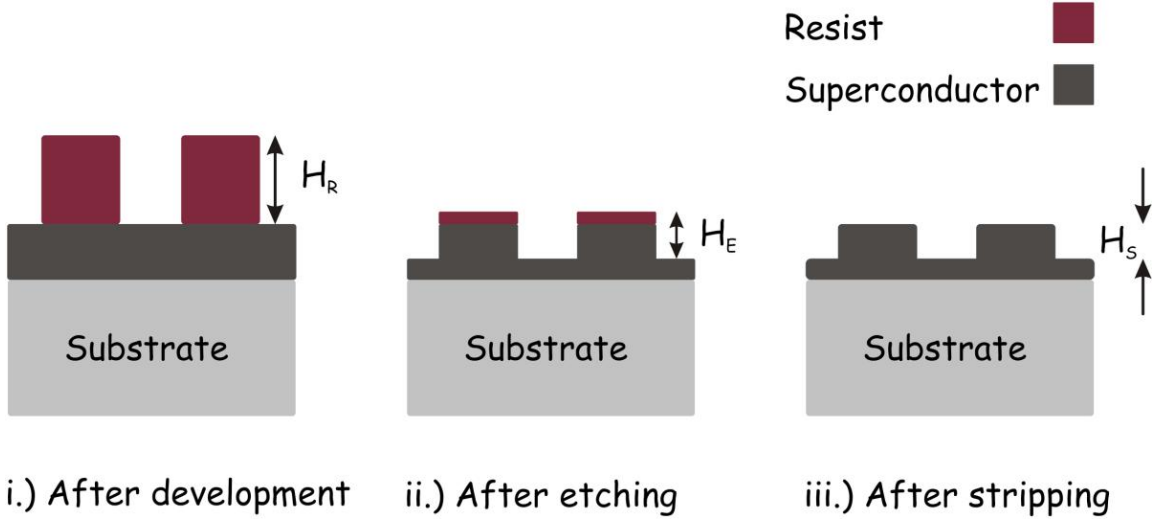

Figure 2: For determining the etching rate $\gamma$, the heights were measured either with an AFM or a standard surface profiler after the development, after the etching and after removal of the resist, respectively. Refer also to figure 5 where photolithographically generated, cleaved resist structures are shown.

From these values and the time $t_{\text{etch}}$ during which the sample was exposed to the ignited discharge, one can calculate the etching rate of the resist,

$$\gamma_R = \frac{H_R - (H_E - H_S)}{t_{\text{etch}}}, \tag{1}$$

and that of the superconductor,

$$\gamma_S = \frac{H_S}{t_{\text{etch}}}. \tag{2}$$

In an earlier work (see figure 1A in [31]), we demonstrated that the critical dimensions of an EBL approach can be reduced by decreasing the height of the spin-coated electron-sensitive resist. For this project, we used PMMA 950k [63]. We explained the observed increase in resolution by a reduced forward-scattering spatial length scale for thinner resists and thus were able to define metallic line-widths in the sub$-15$nm regime. Making use of this approach for fabricating nanostructures within an etching approach means that one has to search for a chemically highly stable resist which survives for a long time in the reactive discharge to allow for the definition of a thin etch mask and hence for a resulting very narrow conduction path width $w$ (see figures 1 and 7).

Therefore we investigated the etching rates for PMMA 950k as well as ZEP 520A in chemically different discharges. The common plasma chemistry was fluoride-based because it is well known that Nb and other transition metals as well as their nitrides are particularly reactive to this type of chemistry [64]. In order to explore the strength of the impact of the fluoride chemistry, the etching rates were measured for four different plasmas ($SF_6$/Ar 10 sccm/30 sccm; $CF_4$/Ar 10 sccm/30 sccm; $CHF_3$/Ar 10 sccm/30 sccm and Ar 30 sccm). Pure Ar was used to additionally determine the rate of pure physical sputtering. The chemically active species made up only 25% of the discharge volume to obtain a relatively low etching rate that is appropriate for the controllable structuring of 5 nm thin films for detector applications. The etching rates at higher DC-bias voltages were investigated to determine how the rates behave with increasing acceleration-voltage of the ions towards the surface. The corresponding results for the two investigated electron-sensitive resists are displayed in figure 3.

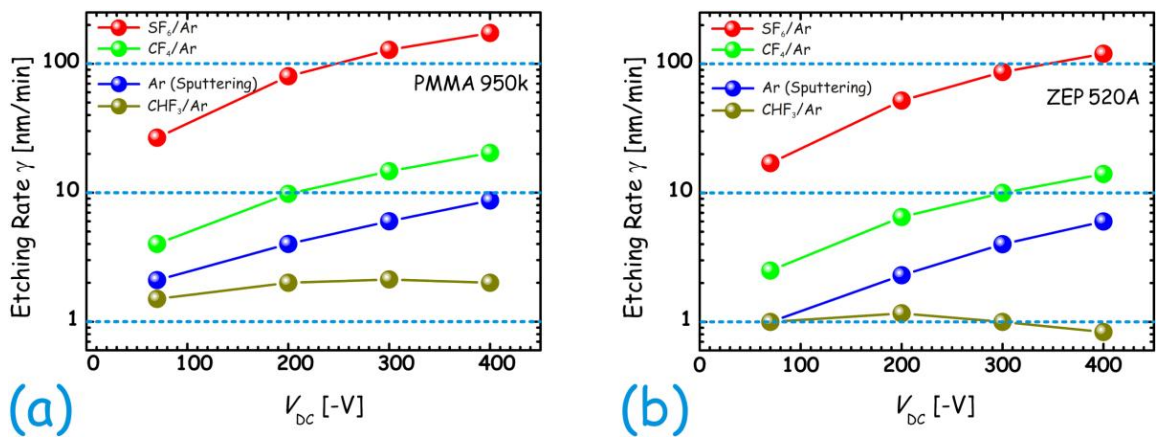

Figure 3: Etching and sputtering rates plotted on a logarithmic scale for PMMA 950k (a) and ZEP 520A (b) as a function of the sheath voltage $V_{DC}$. The blue dotted lines are guides to the eye for the decadal units on the ordinate. The graphs demonstrate that ZEP 520A is chemically more stable than PMMA 950k, resulting in lower etching rates and therefore rendering it the resist of choice for a subtractive lithographic pattern transfer in the sub−100nm regime.

It turned out that ZEP 520A is chemically more robust than PMMA 950k in a number of fluoride-based plasmas (compare figures 3(a) and (b) and consult table 1 for tabulated values at $V_{DC} = -70V$). A comparison of the absolute numbers yields a roughly 1.5 times higher chemical stability of the ZEP 520A for all four different discharge chemistries. Consequently, the thickness of ZEP 520A can be reduced well below $H_R = 100$nm, for detector fabrication out of nanometer-thin films, allowing therefore for the fabrication of sub−100nm patterns (according to figure 1A in [31]) and assuring at the same time the complete protection of the underlying structure during the dry etching. The data of figure 3 served as the basis for our decision to use ZEP 520A as the etch protection layer for the fabrication of the nanoscaled NbN-based superconductors as the ones shown in figures 1 and 7.

As first sight it seems that the etching rate for the $CHF_3$/Ar-plasma discharge decreases with increasing sheath voltage (see figure 3(b)). However the absolute etching-rates $\gamma$ for this discharge-operation vary only in the interval $[0.83-1.17\text{nm/min}]$, and therefore this "effect" might have its origin in experimental uncertainties during the measurements of the resist heights ($t_{etch} = 20$min in this case). Nevertheless, it is particularly interesting that for both organic resists the etching rates in the $CHF_3$/Ar-plasma are lower than for the pure physical sputtering within the Ar-discharge. We note that the required kinetic energy of the charged ionic particle accelerated through the plasma sheath (see chapter 3.1) which induces the ablation-reaction on the sample's surface is too low in our $CHF_3$/Ar-plasma.

The etching rates for different low-temperature superconducting metals were finally investigated only in the SF$_6$/Ar discharge at $V_{\text{DC}} \approx -70\text{V}$. This relatively low ion-acceleration voltage was chosen to minimize damages to the edges of the conduction path during the etching and therefore to minimize the resulting line-edge roughness [35]. In addition, the resulting low etching rate allows for a controlled material-ablation with high precision which is highly desirable for such nanometer-thin films. We also tried to determine the etching and sputtering rates for the superconductors in the other three above described plasmas. However, at $V_{\text{DC}} = -70\text{V}$, no measurable surface abrasion after the exposure in the discharge could be observed. Because only the SF$_6$/Ar discharge is suitable for a subtractive lithographic structuring of our superconducting films, the AFM-obtained etching rates (see figure 4(a)) for this particular type of plasma discharge is brought in the following into focus. The etching rates for some materials were determined for slightly different sheath-voltages which absolute values are slightly higher than $V_{\text{DC}} = -70\text{V}$. The Nb-based superconductors show the highest robustness against the reactive dry-etching mechanism that is accompanied with the lowest etching rate determined during our investigation $(\gamma \approx 2\,\text{nm/min})$. The MoN and the TiN have approximately the same etching rate $(\gamma \approx 9\,\text{nm/min})$ while TaN is chemically more robust $(\gamma \approx 6\,\text{nm/min})$ than pure elemental tantalum $(\gamma \approx 14\,\text{nm/min})$.

In addition we investigated the etching rates for both silicon and sapphire substrates. The sapphire was not affected by the discharge, while the silicon showed a relatively high etching rate in the SF$_6$/Ar plasma as already mentioned in chapter 2. Because our complete lithographic process mixes and matches EBL and photolithographic fabrication procedures (for a detailed discussion of our used lithographic step order refer to chapter 4 in [20]), we also investigated the etching rate of the positive organic photosensitive resist AZ6632.

The thus obtained etching rates of the different materials in the SF$_6$/Ar discharge are shown in figure 4(a). The etching rates for ionic energies $<100$eV are also tabulated in table 1. For comparison the physical sputtering rates for the organic resists and silicon are shown in figure 4(b). The sputtering rates are about one order of magnitude smaller than the etching rates, which is in agreement with an intuitive expectation. Among the materials compared, silicon shows the most pronounced reaction to the fluoride-based chemistry's influence (see figure 4).

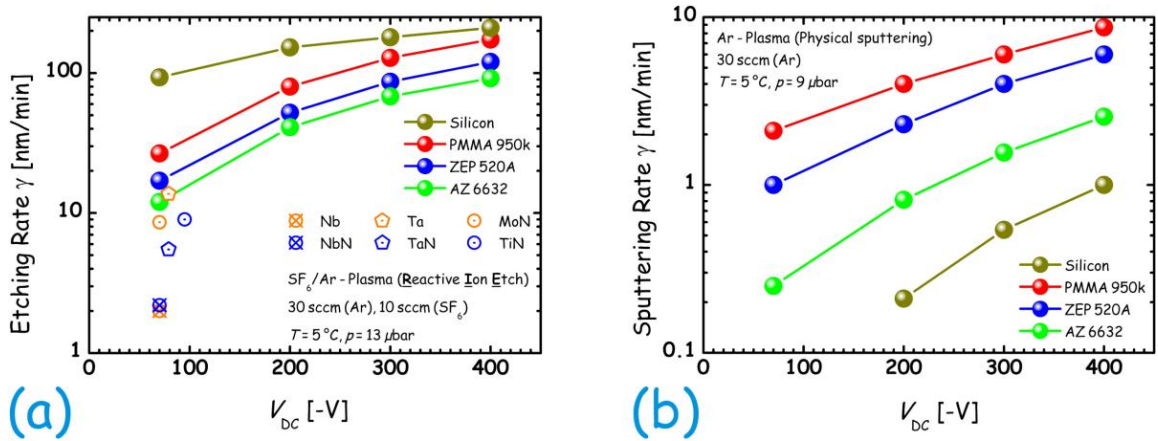

Figure 4: Graph (a) shows the etching rates for the organic resists, silicon and the superconductors in the SF$_6$/Ar discharge which has the highest chemical activity. Graph (b) displays the pure physical sputtering rates for the resists and silicon. The etching rate of the sapphire is negligible.

| $V_{DC}$ | $\gamma_{Nb}$ | $\gamma_{NbN}$ | $\gamma_{Ta}$ | $\gamma_{TaN}$ | $\gamma_{MoN}$ | $\gamma_{TiN}$ | $\gamma_{Si}$ | $\gamma_{PMMA\,950k}$ | $\gamma_{ZEP\,520A}$ | $\gamma_{AZ\,6632}$ |
|---|---|---|---|---|---|---|---|---|---|---|
| -70 V | 2.0 nm/min | 2.2 nm/min | - | - | 8.6 nm/min | - | 93.2 nm/min | 27.0 nm/min | 17.0 nm/min | 12.0 nm/min |
| -79 V | - | - | 13.7 nm/min | 5.5 nm/min | - | - | - | - | - | - |
| -95 V | - | - | - | - | - | 9.0 nm/min | - | - | - | - |

Table 1: Low-power etching rates for six different low-temperature superconducting metals, silicon and three organic photon- and/or electron-sensitive resists (see also figure 4(a)).

Finally, to observe directly the quality and the shape of the profile of the pattern transfer generated by our subtractive reactive dry-etching approach, we performed contact photolithography with the positive AZ6632 resist (from *MicroChemicals*) that we spin-coated to a height $H_R = 2.10\mu m$ on top of a silicon wafer (the silicon was used because it allows for cleaving the wafer). We employed our previously published approach (see chapter 3 in [20] where the procedure is explained in more detail) that allows for the extension of the spatial fabrication scale of a relatively low-cost optical contact photolithographic approach with a mercury arc-discharge photon source down to the order of the wavelength ($\text{sub}-\mu\text{m}$ in case of a mercury discharge) of the used light. To resolve these spatial frontiers of contact photolithography, a matrix of fine straight lines with different widths and spacings is included on our optical chromium mask (see figure 8 and *"Cleaving structures"* in figure 9 of reference [20]). In the horizontal direction the line spacing is varied, while in the vertical direction the line width is varied (both varied five times: $0.5\mu\text{m}$, $1\mu\text{m}$, $2\mu\text{m}$, $3\mu\text{m}$ and $4\mu\text{m}$). Figure 5(a) shows the result for $3\mu\text{m}$ wide lines spaced $1\mu\text{m}$ after the chemical development (in pure AZ 726 MIF from *MicroChemicals*) at room temperature for $1\text{min}$. The exposure dose $\left(25\text{mJ/cm}^2\right)$ was adjusted with respect to the Hg i-line $\left(\lambda \cong 365\text{nm}\right)$. We used half of the value that was recommended by the manufacturer to obtain a longer development time and therefore a better control of the chemical developing process. Afterwards we performed a cleaving procedure on a set of lines that were $1\mu\text{m}$ wide and spaced $2\mu\text{m}$ apart from each other (see figure 5(b)) and investigated the profile of the resist with a scanning electron microscope (SEM). The chemical development of the AZ6632 yielded a trapezoid profile with an angle $\varphi \cong 80°$ with respect to the sample surface (defined by the right-angled triangle in figure 5(b)).

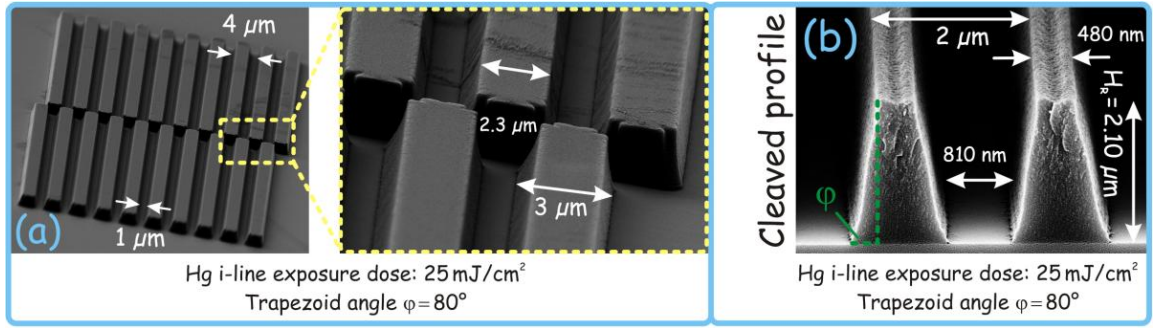

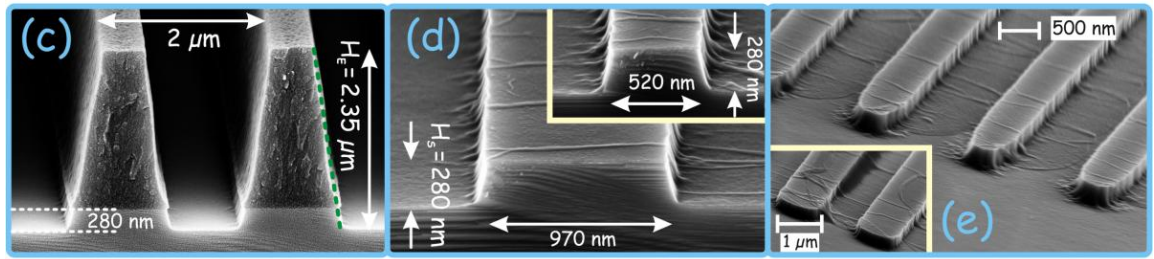

Figure 5: The electrographs (a) show the unexposed parts of the positive photoresist AZ6632 at two different magnifications after the chemical development. The picture (b) shows the cleaved profile of this photoresist that is generated during the development. Picture (c) was generated after the sample was exposed to the $SF_6$/Ar discharge. The green dotted line is a guide to the eye to illustrate that the trapezoid shape of the resist (b) is transferred subtractively into the wafer surface by the ionic bombardment (see chapter 3.1). Electrographs (d) and (e) were scanned after the resist residuals were stripped away in hot NMP and show the critical dimensions of our dry etching lithographic process that are as narrow as the ones in figure 7(f) in [20]. All structures are located on a silicon substrate to allow for a cleaving inspection with the SEM. Details are explained in the main text.

Another chip was lithographically prepared in the same manner and exposed afterwards to the reactive $SF_6$/Ar discharge for $t_{etch} = 3\,\text{min}$. After the etching we repeated the cleaving procedure and measured $H_E \cong 2.35\,\mu m$ as shown in figure 5(c). A very important information contained in figure 5(c) is that even at the lowest investigated sheath voltage $V_{DC} = -70\,V$ (chosen to ensure a sensitive thin-film removal), the trapezoid shaped profile generated during the chemical development during photolithography is transferred by the reactive plasma discharge into the silicon wafer surface (see green line in figure 5(c)). This is a strong evidence for a nearly complete directional etching, hence justifying our low-pressure considerations of chapter 3.1. We note that a slight under-etching due to the isotropic character of the chemical reaction is also recognizable in figure 5(c). However, the vertical component of the reaction induced by the physical acceleration of the ions through the sheath is much larger.

Finally, we processed a third chip where the resist was stripped away after the dry etching in 85°C hot NMP, and we electrographed the critical dimensions of our contact photolithographic approach in the spatial cross-over regime from the microscale into the nanoscale (see figures 5(d) and 5(e)). We note that our subtractive dry-etched photolithographically generated pattern yields the same spatial resolution (close to the wavelength of the mercury arc discharge lamp, $\lambda = 365 \text{nm}$) as the one obtained using an additive non-invasive lift-off procedure (see figure 7 in [20]). From the measured values $H_R, H_E, H_S$ shown in figure 5, we could reproduce (using equations (1) and (2)) the etching rate for silicon and the AZ6632 as determined with the AFM-method (figures 2 and 4 and table 1).

**4. Electron-Beam Patterning**

Considering the results discussed in the previous chapter (see table 1), the preference of ZEP 520A as etch protection mask in EBL is obvious, because it allows for the definition of smaller critical dimensions than PMMA 950k according to figure 1A in [31]. From the determined etching rates listed in table 1 we can conclude that a 5 nm thick NbN film needs a $\cong 70 \text{nm}$ thick ZEP 520A etch protection to guarantee an intact NbN nanoscaled structure after the lithographic etching (we etch $t_{\text{etch}} = 3 \text{min}$ to ensure that all of the unprotected NbN has been removed; then there are still $\approx 21$nm ZEP 520A left that guarantee the protection of the photon detector structure). Therefore, the EBL investigations described in this chapter were performed with this particular thickness of ZEP 520A which determines the obtainable critical dimensionality [31]. We used a Raith150 EBL tool that we operated at its maximum electron acceleration voltage of 30kV to minimize the forward scattering scale and thus increasing the lithographic resolution.

From a fabrication engineer's point of view, such a high tension is unfortunately accompanied by an unwanted increase of the probability for elastic large-angle backscattering events (see chapter 4.1 in [65]) of electrons that come close to the atomic nuclei of the substrate elements. This backscattering process is characteristic for a given substrate material and strongly depends on the density and the number of protons of the atoms forming the substrate. This unintended exposure generated by backscattered electrons is called *electron-beam lithographic proximity effect* [66-68]. Due to the positive character of the ZEP 520A, the electron-beam has to be guided around the conduction paths of the photon detector to generate the etch mask during the chemical development. This exposure around the desired structure makes corrections for the proximity effect almost redundant for nanowire meander-detector device layouts (see chapter 2 in [31] for a detailed discussion). Therefore, the EBL-fabricated structures of figures 1, 6 and 7 were written by the scanning electron-beam without correcting for proximity exposures.

The clearing dose of an electron-sensitive resist is defined as the sum of forward- and backward-scattered electrons per unit area. To determine this dose for ZEP 520A, relatively large squares $(25 \times 25 \mu m^2)$ which are in both planar dimensions much larger than the backscattering length ($2.436 \mu m$ at 30kV for sapphire after [69]) were exposed. It is well known that ZEP 520A has a higher intrinsic contrast as compared to PMMA 950k [36]. The resulting image-replication accuracy during the lithographic pattern transfer is therefore higher when using the ZEP 520A. However, the intrinsic contrast of an organic resist increases by chemically developing it at lower temperatures [36]. The higher contrast makes the ZEP 520A even more robust against proximity exposures, and it also increases the lithographic line-edge roughness [35]. Consequently, the resulting clearing dose was determined to $55\mu C/cm^2$ at $-10°C$ developer temperature (the value at room temperature is $45\mu C/cm^2$).

These relatively low values for the clearing dose result in high beam-speeds $v_B$ that are accompanied by pronounced dynamic effects during the acceleration and deceleration of the deflected scanning-beam that originate from the finite inductance of the deflection coil system of the lithograph. These dynamic effects, if not suppressed properly, can cause significant vacancies in the exposed pattern. The lowest possible beam-current of $27 pA$, determined by the smallest aperture of $10\mu m$ in diameter, and the used scanning step size of $10 nm$ result in a beam-speed of about $5 mm/s$ for depositing the above determined clearing dose at $-10°C$. Because of the higher clearing dose, the lower developer temperature diminishes the dynamic effects in addition to its contrast enhancement. Under these conditions and without activating the dynamic compensation module implemented within the Raith software suite, we still observed dynamic effects for lateral fabrication dimensions in the $sub-150 nm$ regime (not shown). However, with the dynamic compensation module activated, pattern on the $sub-50 nm$ scale can be generated with very high accuracy (see figures 1, 6 and 7).

After the clearing dose was determined at $-10°C$ and the dynamic effects suppressed, the critical lithographic dimensions were investigated by taking into account the design of the single-photon detectors which has basically three planar design parameters. The width of the nanowires $w$, their spacing $s$ and the detection area $A_{detec}$ (see figure 1). Consequently, the parameters $w$ and $s$ were varied to allow for the investigation of the critical dimensions of isolated lines $(s \gg w)$ and their critical spacing $(s \ll w)$. With this approach, the smallest critical conduction path width $w$ and the highest possible filling factor of the latter detector (determined by $s$) that are allowed by the lithographic processing could be deduced. Among other things [43], the filling factor ultimately determines the quantum efficiency of the operational photon detector (see chapter 13 in [18]). The design width of the conduction paths $w$ within the GDSII-editor (Graphic Data System) of the Raith software suite was varied from 300 nm to 20 nm (300 nm, 200 nm, 160 nm, 120 nm, and from 120 nm to 20 nm in 10 nm steps). For $w \geq 200$ nm the detection area covered about $10 \times 10\,\mu m^2$. For smaller conduction paths $w \leq 200$ nm the corresponding area was about $5 \times 5\,\mu m^2$.

In the direction parallel to the current path (see figure 1), the detection area length was exactly $10\,\mu m$ (or $5\,\mu m$) while in the direction perpendicular to it the number of meander turns was chosen in a way that the detection area width was slightly less than $10\,\mu m$ (or $5\,\mu m$). This was done in order to reduce the kinetic inductance of the superconducting detector and therefore the pulse duration (according to [70]), but still maintaining an approximately square geometry at the same time. For each conduction path width $w$ three meanders with different $s$ values were designed in order to obtain three filling factors (0.25, 0.5 and 0.75). This whole entity of 45 EBL-design devices allowed us to explore *design* widths $w$ down to 20 nm and *design* spacing $s$ down to about 7 nm.

We found a resulting critical design dimension $w = 90$ nm and a critical design spacing of $s = 60$ nm for the 70 nm thick ZEP 520A. A smaller $w$ resulted in unphysical and irreproducibly erratic resistance curves, in very contrast to the smooth curves shown in figure 8 for $w = 90$ nm (we will see in the following that a designed width of 90 nm results in a 53 nm wide superconducting NbN-path as the one shown in figures 1 and 7). A smaller $s$, on the other hand, leads to merging of the conduction paths after the plasma etching (not shown). Both values can be, in principle, reduced for thinner resists (according to figure 1A in [31]). Therefore we suggest the usage of a chemically more stable etch protection mask than ZEP 520A for lithographically fabri-cating even narrower superconducting paths, and we note that the chemical robustness of the ZEP 520A can indeed be increased as shown by [71].

Figure 6 displays a section of the structured ZEP 520A for the critical design dimension $w = 90\,\text{nm}$ after the exposure and chemical development for two different temperatures. The lower developing temperature (figure 6(b)) results in a higher contrast, in excellent agreement with [36]. However, the chemical development shrinks the designed value of the current path-width by about $30\,\text{nm}$.

Finally, an investigation of the influence of the scan direction was performed for the scanning step size $5\,\text{nm}$ and $10\,\text{nm}$. We found that the dynamic effects are more pronounced for the smaller step size (and hence higher beam-speed), in excellent agreement with an intuitive expectation. Best results were achieved for guiding the electron-beam along the direction of the current paths (shown in figures 1, 6 and 7; exposure mode analogous to figure 3 in reference [20]). If the beam is guided perpendicular to the current path, the interruption of the scanning beam at the edge of the conduction path is clearly recognizable after the chemical development. This effect becomes more pronounced for patterns in the $\text{sub}-100\,\text{nm}$ regime and for higher scanning speeds (not shown).

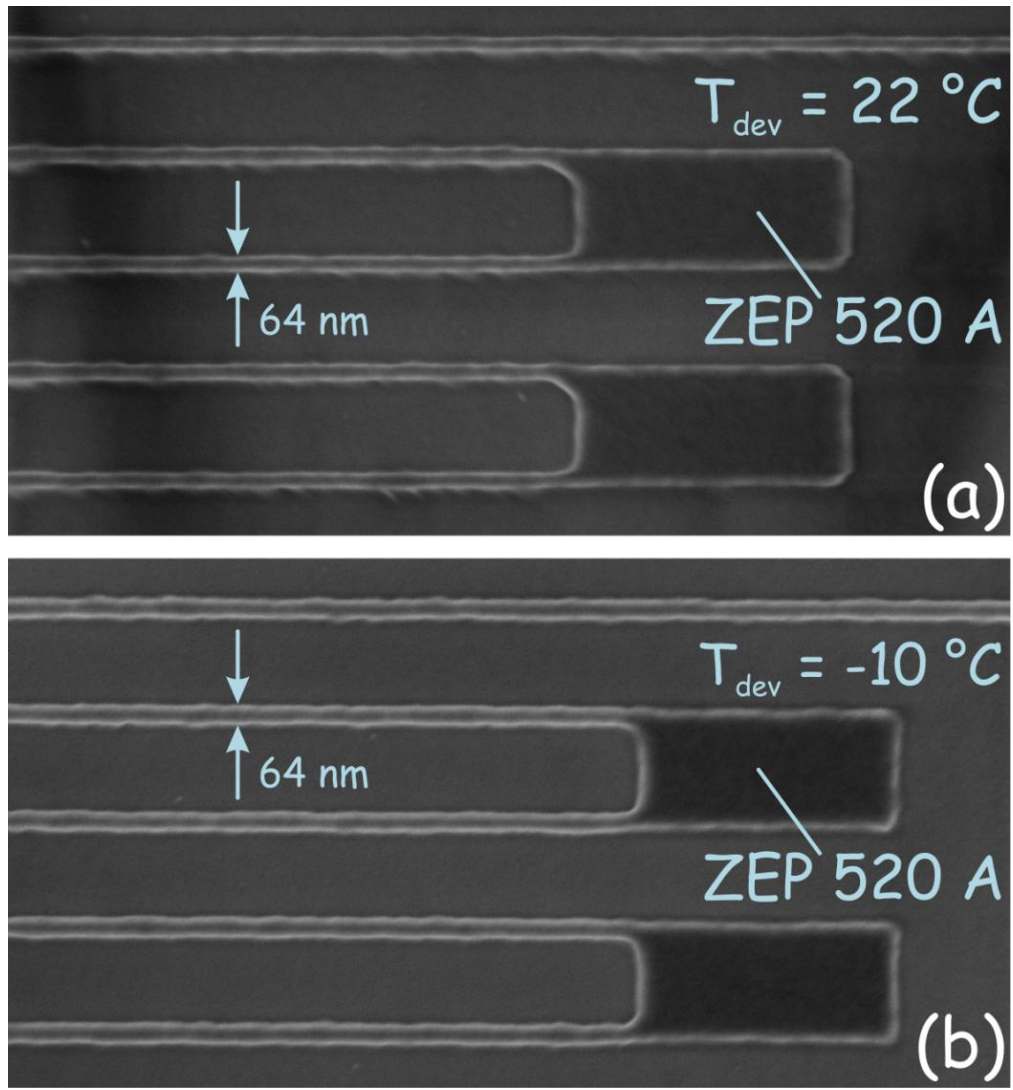

Figure 6: Unexposed ZEP 520A nanoscaled structures after the chemical development in n-Amylacetate. The structure of electrograph (a) was developed at room temperature, while electrograph (b) shows the resulting ZEP 520A after the development at $-10°C$. The low-temperature development leads to a higher contrast, in excellent agreement with [36]. The exposed structures were designed with $w=90$nm and a filling factor $FF=0.25$. After the development $w$ is reduced to $\cong 64$nm, and the corresponding filling factor also decreases as compared to the designed value.

## 5. Results

*5.1 SEM-Inspection after the Etching*

To determine the final geometrical dimensions of the fabricated meander structures, the above mentioned entity of 45 EBL-design devices was lithographically dry-etched 8 nm into a 10 nm NbN film (the height of the etch mask ZEP 520A was $\cong 70$ nm as in the last chapter). The remaining 2 nm NbN ensures the drain of the electrons from the scanning beam during electrography, allowing therefore for the investigation of the meanders with the SEM. In figure 7, an example of the explained technique after etching and stripping of the resist is shown (see also figure 1).

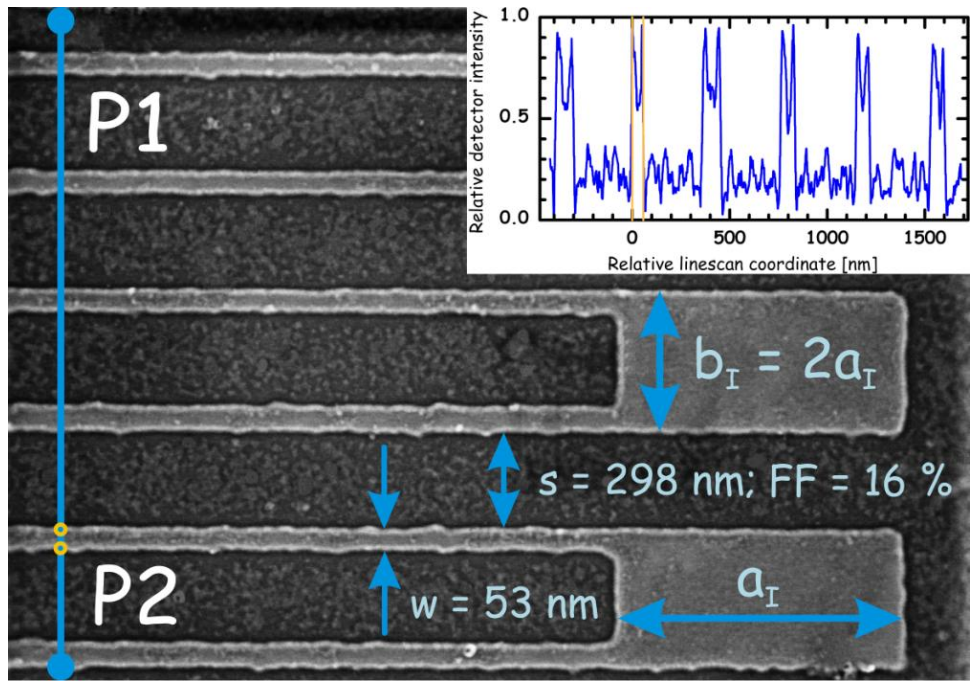

Figure 7: The electrograph shows the etched nanoscaled meander that was protected by the ZEP 520A structure shown in figure 6(b). The upper inset shows a line-scan of the scanning beam guided along the blue line. The structure was not corrected for proximity exposures during the electron-beam lithography (according to chapter 2 in [20]).

Like the chemical development, the etching further reduces the designed conduction path width (by about 10nm, compare figures 6(b) and 7). The discrepancy for the width $w$ of about 30nm between the designed value and the SEM measured values was indeed observed on every structure within the entity of 45 EBL-design devices. Measuring the width of the conduction path nine times within the detector according to figure 7, we determined $w=53.4\text{nm}\pm2.3\text{nm}$. This accuracy was obtained without an elaborated correction for the proximity effect during the EBL, and is competitive to [69], where the proximity effect was corrected during the fabrication procedure (refer to chapter 2 in [20] for a more detailed discussion).

*5.2 Electronic Transport Measurements of the Fabricated Photondetector Structures*

The transport measurements were done at low bias current $I_b = 500\text{nA}$ in a Physical Property Measurement System (PPMS from *Quantum Design*) in various magnetic fields $B$ up to 9T perpendicular to the thin film surface. In the following we focus on the resistance properties of a device clone of the structure electrographed in figure 7.

The temperature dependent specific resistance $\rho(T)$ was calculated as

$$\rho(T) = R(T) \cdot \left(\frac{L}{w} + \frac{N}{2}\right)^{-1} \cdot d \tag{3}$$

with $R$ the measured resistance ($R(22\text{K}) \cong 0.67\text{M}\Omega$) value, $L = 73.9\mu\text{m}$ the entire length of the nanowire (without the islands) and $N = 12$ the number of islands of dimensions $a_I \cdot b_I$ connecting the strips ($N = N_p - 1$, with $N_p$ the number of conduction paths). The factor 2 in equation (3) stems from the design rule $b_I = 2a_I$. The islands were placed to avoid a possible constriction when the conduction path turns around by 180°. Such a constriction results in a reduced critical-current which might limit the current through the whole device and therefore reduces the sensitiv-ity of the detector [16,17]. If the islands are not included $(N=0)$, the resistance value defined by equation (3) is altered within an uncertainty of $\approx 1\%$.

After film growth and nanopatterning, the exposure of the NbN to air leads to an oxidation of surface and edge layers and therefore to a suppression of superconductivity [72,33] within these layers, which will influence the superconducting core of the conduction paths via the superconducting proximity effect [34,72]. Therefore, reduced geometrical dimensions were used for the quantitative analysis of the experimental data, i. e. $5\,\text{nm}$ from the width $w = 53\,\text{nm}$ as determined with the SEM (see figure 7) and $1\,\text{nm}$ from the height $d$ as determined with the AFM were subtracted (refer to chapter II. A. of [19]). The thus computed resistivity data are displayed in figure 8(a).

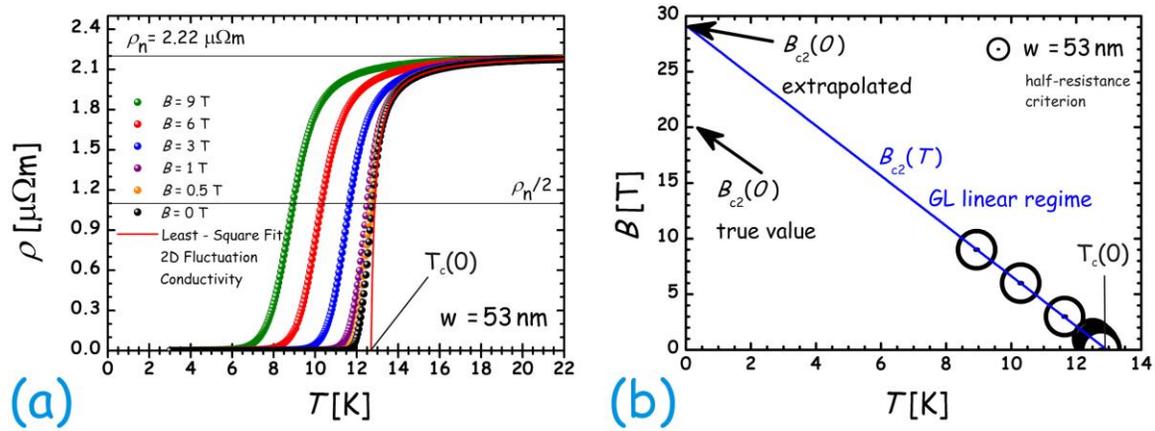

Figure 8: Graph (a) shows the electronically measured phase transition that separates the metallic from the superconducting state for different magnetic fields up to $9\,\text{T}$ for a device clone of the sample shown in figure 7. The red curve is a least-squares fit according to equation (4) which is based on two-dimensional fluctuation conductivity [73]. Graph (b) shows the linear extrapolation of the upper-critical field $B_{c2}(T)$. The experimentally determined $B_{c2}(T)$ values were extracted from graph (a) by a half resistance criterion. The extrapolated value at $T = 0\,\text{K}$ was corrected by $69\%$ to obtain the true value $B_{c2}(0)$ (according to [74]).

Close to the transition into the superconducting state at the critical temperature $T_c(B)$, thermal fluctuations dominate the physics of the device, leading to a rounding of temperature dependent resistance, which is more pronounced in lower dimensional systems [25,75]. These fluctuations on the order of $k_B T$ generate short-lived Cooper-pairs [76] which enhance the conductivity of the sample even relatively high above $T_c(B)$. A precise analysis of this rounding of the electronically measured resistance data in zero field thus allows for drawing conclusions about the dimensionality of the fabricated superconducting structure [25,75]. We achieved best results by least-squares fitting the experimental data of $\rho(T)$ for $B=0\,\mathrm{T}$ to a Cooper-pair fluctuation model developed for two-dimensional superconducting systems [73] (see red line in figure 8(a))

$$\rho(T) = \frac{\rho_n}{1 + \rho_n \cdot C \cdot \frac{1}{16} \frac{e^2}{\hbar d} \frac{1}{t-1}}, \qquad t = \frac{T}{T_c(0)}, \qquad T > T_c(0). \qquad (4)$$

Here $\hbar$ is the reduced Planck's constant, $e$ is the elementary charge, and $C$ is a fitting parameter with $C=1$ in an ideal two-dimensional system (we obtained $C=1.8$). This procedure allowed us to analytically extract the critical phase transition temperature in zero magnetic field $T_c(0)$ and the normal-state resistance $\rho_n$ (see figure 8(a)).

The critical temperatures $T_c(B)$ and the upper critical fields $B_{c2}(T)$ were determined using a 50% resistance criterion. This criterion allowed us to determine the $T$-dependent magnetically induced phase transition line $B_{c2}(T) = \mu_0 H_{c2}(T)$, with $\mu_0$ the permeability of vacuum. As shown in figure 8(b), the upper critical field is, to a very good approximation, linear in temperature up to the highest experimentally available magnetic field $B=9\,\mathrm{T}$, as expected from Ginzburg-Landau (GL) theory [77] in the vicinity of $T_c(0)$.

Within the framework of the GL theory, $B_{c2}(T)$ is related to the quantum of the magnetic flux, $\Phi_0 = h/2e \approx 2.07 \cdot 10^{-15}$ Vs and to the temperature dependent coherence length $\xi(T)$ by

$$B_{c2}(T) = \frac{\Phi_0}{2\pi\, \xi(T)^2} \;. \tag{5}$$

From a linear extrapolation of $B_{c2}(T)$ to zero temperature (see figure 8(b)) we obtain the true value for the upper critical field $B_{c2}(0)$ by multiplying the extrapolated value with 0.69 (after Werthamer *et al.* [73]). This approach leads to $\xi(0) \cong 4\,\text{nm}$, justifying again the two-dimensional character of the 5 nm thick structure. For a more profound analysis of the measured data in terms of the superconducting parameters we refer to chapter II. B. in [19].

Superconducting meanders fabricated with the approach discussed in this paper have served as excellent detectors for thermally fluctuating magnetic flux quanta [19] and for single photons (see chapter 13 in [18]) in the visible and near-infrared range with a pulse duration of $\sim 6\,\text{ns}$ nanoseconds. Our detectors registered visible photons travelling through the detection area with an efficiency of $\sim 6\%$ that is mainly determined by the filling factor $FF$. The critical temperatures of 16 nanoscaled samples (fabricated with the top-down approach shown in figure 9 of [20]; the nanolithograph approach was described above) with different planar geometric dimensions (see tables 11.1-11.4 in [18]), characterized by fitting equation (4) to the measured resistance data varied only within $T_c(0) = 12.63\,\text{K} \pm 0.07\,\text{K}$ and the normal state resistances within $\rho_n = 2.1\,\mu\Omega\text{m} \pm 0.12\,\mu\Omega\text{m}$. This underlines an excellent repeatability and yield of our optimized lithographic planar top-down production approach [20]. The critical temperature of the nanostructures was reduced by only about $0.5\,\text{K}$ as compared to the $T_c(0) \cong 13.2\,\text{K}$ of the unstructured film, indicating negligible damage and chemical modification which is accompanied by a very weakly pronounced superconducting proximity effect. The films for our 16 samples were in addi-tion sputtered in different fabrication runs, indicating the optimized interplay between our sputtering and nanoscale fabrication approach. Our fabrication approach and its results are directly comparable to the best results reported by other research groups [78-83].

Finally, the most convincing argument for the excellent quality and uniformity of the conduction paths of our superconducting meander structures (besides the lithographic arguments shown in figures 5(c), 6(b) and 7) is their ability to carry bias currents up to 60% of the theoretical de-pairing-current limit (see table 1 and equation (10) in [19]). Already a small number of constrictions along the total length of the meander would significantly limit the experimental critical current whereas the material parameters that determine via equation (10) in reference [19] the theoretical de-pairing current-limit are obtained from low-current resistivity measurements and would not be noticeably affected by small variations of the cross-sectional area.

## 6. Conclusion

We have developed a planar top-down $2\frac{1}{2}$D fabrication approach for defining high quality photon-sensitive superconducting current-paths on the 50nm scale that are able to permanently carry critical currents up to 60% of the theoretical de-pairing current-limit. Furthermore, we optimized the DC film-sputtering technique to obtain high transition temperatures even for 5nm thin NbN films. In addition, the material ablation on the nanoscale in a reactive RF-plasma was optimized in a symbiotic manner with the EBL, in order to minimize the line-edge roughness and to retain the high phase-transition temperature after the nanoscale lithography. From measurements of the etching rates we conclude that ZEP 520A as etch protection mask allows for the fabrication of narrower structures as compared to PMMA 950k because of its higher chemical stability. The developed etching recipe turned out to be appropriate to structure a number of metals, organic resists and silicon. Our lithographic approach might be useful and applicable for other re-search tasks such as field-effect transistors [84], graphene-electronic components [85] or single-electron devices [86], where the controlled narrowing of thin films of a few atomic monolayers on the $\text{sub-100 nm}$ scale is desirable.

Finally, the fabricated superconducting devices were characterized in the framework of the fluctuation conductivity theory and the standard GL-theory. The value obtained for the coherence length from the GL theory evidences the two-dimensional character of our devices, which is in agreement with the outcome of the fluctuation conductivity analysis. The successful least-squares fitting of the results of the seminal and mathematically challenging theories to the experimentally measured resistivity data of our devices emphasizes the optimized interplay between our film generation and our film structuring techniques and it justifies the high quality of our devices.

**Acknowledgements**

This work was partially supported by the Swiss National Science Foundation program NCCR Materials with Novel Electronic Properties (MaNEP) and by the DFG Center for Fuctional Nanostructures. HB acknowledges the Research Funding 2010 from the University of Zürich. The TiN film was supplied by D. Jäger, EMPA, Switzerland. Devices were fabricated at the FIRST Center for Micro- and Nanoscience of the ETH Zürich. We are thankful to G. Piaszenski and the team from Raith GmbH in Dortmund for a successful and fruitful collaboration. Throughout this paper the international SI unit system was used.